Transformation Journey of Zr-based MOFs: Study on Mechanics and Hydrogen Storage under Doping Regulation

Yanhuai Ding[*], Dan Qian, Zhipeng Liu

School of Mechanical Engineering and Mechanics, Xiangtan University, Xiangtan 411105, China

[*]Corresponding author. E-mail: yhding@xtu.edu.cn

**Abstract**

This study delves into the transformation journey of Zr-based Metal-Organic Frameworks (MOFs), focusing on enhancing their mechanical properties and hydrogen storage capacities through doping regulation. MOFs, a versatile class of crystalline porous materials, have garnered significant attention due to their unique properties and broad potential applications in gas storage, separation, catalysis, and sensing. Among them, Zr-based MOFs stand out for their exceptional stability and high surface area. This research systematically investigates six key Zr-based MOFs (UIO-66, UIO-67, UIO-68, MOF-801, MOF-802, and MOF-841) using multiscale computational methods, including molecular dynamics (MD) simulations, grand canonical Monte Carlo (GCMC) simulations, and density functional theory (DFT). The study explores the impact of metal ion substitution (Fe, Co, Ni, Cu, Zn) on the mechanical and hydrogen storage properties of these MOFs. Our findings reveal that metal ion substitution significantly influences the mechanical stability and hydrogen adsorption capacity of Zr-based MOFs, providing valuable insights for the design and optimization of high-performance MOF materials.

# 1. Introduction

Metal-organic frameworks (MOFs) have garnered significant attention as a versatile class of crystalline porous materials, owing to their unique properties and broad potential applications[1-3]. MOFs are constructed through the coordination of metal ions or clusters with organic linkers, resulting in highly ordered and tunable structures[4, 5]. Since their initial introduction in the late 20th century, substantial progress has been achieved in the synthesis, design, and application of MOFs. The synthesis of MOFs can be realized through various routes, including hydrothermal and solvothermal methods[6]. These methods enable the creation of a diverse array of topologies, morphologies, and composites, highlighting the adaptability of MOFs. The tunability of MOFs' structures and functions through linker design is a key feature that has driven their widespread application in fields such as gas storage, separation, catalysis, and sensing[7, 8]. For instance, MOFs have demonstrated exceptional performance in hydrogen storage and carbon dioxide capture. Their porous nature and high surface area also make them promising candidates for environmental remediation and energy storage applications. Moreover, MOFs have shown great potential in biomedical applications, such as drug delivery and bioimaging. However, despite their numerous advantages, MOFs still face challenges, including issues related to stability and scalability. Ongoing research is focused on addressing these challenges and further exploring the potential of MOFs in various fields[9, 10]. As a result, MOFs continue to be a vibrant area of research, with the potential to address a wide range of global challenges.

After years of development, MOF materials have formed a vast family. Thanks to the tunability of their structures, the variety of MOF materials continues to expand. Among the extensive family of MOF materials, Zr-based MOFs have garnered widespread attention from the academic community in recent years[11]. Zr-based MOFs were initially discovered by Lillerud and co-workers and have since been demonstrated to possess superior stability[12, 13]. The prototypical framework, UiO-66, is composed of inorganic building units with the formula $[Zr_6O_4(OH)_4)]^{12-}$ and terephthalic acid dianions as linkers. These components together form a cubic close-packed topology. The high degree of twelvefold connectivity, along with the strong coordination bonds formed between the tetravalent Zr ions and the linkers, likely contribute to the enhanced stability of the framework.

Over the past decade, studies have highlighted the significant potential of microporous MOFs with high surface areas and micropore volumes for hydrogen storage applications. However, many MOFs exhibit moisture sensitivity. This sensitivity often results in structural decomposition, which in turn compromises the reproducibility of the materials and reduces their hydrogen adsorption capacity. Interestingly, Zr-based MOFs show high structural resistance against water and external mechanical pressure. For example, Jianwei Ren[14] introduced an optimized synthesis method for Zr-based MOFs that enhances ease of handling and reduces reaction time. The resulting Zr-MOFs exhibit excellent thermal and moisture stability, along with a significantly improved hydrogen storage capacity. Liangzhi Xia utilized the Grand Canonical Monte Carlo (GCMC) method to predict the hydrogen storage characteristics of several Zr-based MOFs, including MOF-801, MOF-802, MOF-808, and MOF-841, at 77 K[15]. The study primarily focused on analyzing the hydrogen adsorption isotherms and isosteric heats of these materials, while also investigating the factors that influence their hydrogen storage performance. Sanjit Nayak[16] reported recently introduced three novel Zr-based MOFs: UBMOF-8, UBMOF-9, and UBMOF-31, with UBMOF-31 being synthesized via a mixed-linker approach. The presence of amino groups in UBMOF-31 significantly enhances its hydrogen adsorption capacity, achieving a loading of 4.9 wt% at 4.6 MPa.

This represents one of the highest hydrogen uptake values reported for Zr-based MOFs to date. These studies have shown that structural modulation is one of the best ways to improve the physical properties of Zr-based MOFs.

In this paper, we concentrate on Zr-based MOFs, aiming to boost their mechanical and hydrogen-storage properties. Our objective is to develop high-performance materials with high stability and efficient hydrogen storage. To this end, we employ multiscale computational methods such as MD simulation, GCMC, and DFT. We conduct a systematic study of six key Zr-based MOFs: UIO-66, UIO-67, UIO-68, MOF-801, MOF-802, and MOF-841. Our research delves into their performance differences and explores how metal-ion substitution affects their properties.

## 2. Models and calculation methods

This paper compares the mechanical and $H_2$-storage properties of several Zr-based MOFs. These materials have similar compositions, all being constructed from $Zr_6O_4(OH)_4(-CO_2)_n$ and different linkers. The structures of $Zr_6O_4(OH)_4(-CO_2)_n$ and the linkers are shown in Figure 1, where red, gray, white, and light blue represent O, C, H, and Zr atoms, respectively. UIO-66, UIO-67, and UIO-68 use terephthalic acid, biphenyl-4, 4'-dicarboxylic acid, and triphenyl-4, 4'-dicarboxylic acid as linkers between $Zr_6O_4(OH)_4(-CO_2)_n$ units. MOF-801, MOF-802, and MOF-841 use trans-J-ene-1, 2-dicarboxylic acid, 2, 3-pyridine-dicarboxylic acid, and 4-(carboxymethyl)-benzene-1-carboxylic acid, respectively. Their structural models and parameters, obtained from the Cambridge Crystallographic Database in the UK, are listed in Table 1.

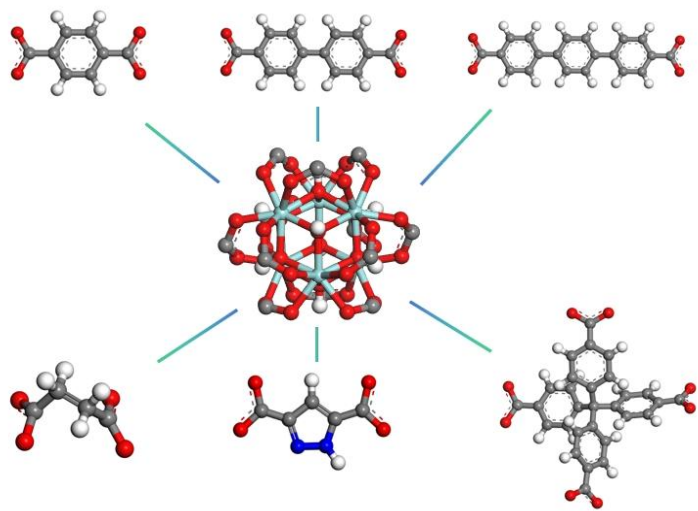

Figure 1 Structures of $Zr_6O_4(OH)_4(-CO_2)_n$ and the linkers

Table 1 Lattice parameters of MOF materials

| No. | a (Å) | b (Å) | c (Å) | α (°) | β (°) | γ (°) |
|---|---|---|---|---|---|---|
| UIO-66 | 20.955 | 20.955 | 20.955 | 90 | 90 | 90 |
| UIO-67 | 26.783 | 26.783 | 26.783 | 90 | 90 | 90 |
| UIO-68 | 32.766 | 32.766 | 32.766 | 90 | 90 | 90 |
| MOF-801 | 17.834 | 17.834 | 17.834 | 90 | 90 | 90 |
| MOF-802 | 39.222 | 26.018 | 27.887 | 90 | 90 | 90 |
| MOF-841 | 14.676 | 14.676 | 28.003 | 90 | 90 | 90 |

The Zr-based MOFs models used in the computations can be constructed in Materials Studio and optimized via its Forcite module with periodic boundary conditions to mimic real-world molecular arrangements. After optimization, the models are treated as rigid frameworks. The L-J potential parameters for framework atoms are from the Dreiding force field. Zr-atom parameters are from the UFF force field, and all specific parameters are listed in Table 2.

Table 2  LJ potential energy parameters of atoms in Zr-based MOFs

|  | MOF_C | MOF_O | MOF_H | MOF_Zr | MOF_N |
|---|---|---|---|---|---|
| $\sigma(\text{Å})$ | 3.473 | 3.033 | 2.846 | 2.783 | 3.263 |
| $\varepsilon/k_B(K)$ | 47.85 | 48.16 | 7.65 | 34.72 | 38.94 |

The mechanical properties of the material were calculated using the Forcite module in Materials Studio. The constant strain mode was chosen, with the number of strain steps set to 4 and the maximum strain amplitude set to 0.003. The COMPASS II force field was selected, and then the model was initiated for the calculation of mechanical properties. Once the software began to run, the strain was applied to the XX, YY, ZZ, YZ, ZX, and XY directions of the model, representing compression and tension in these directions. The bulk modulus, shear modulus, Young's modulus, and Poisson's ratio of the material were calculated based on the obtained elastic stiffness matrix. The calculation of hydrogen storage performance was carried out using the Adsorption module in Materials Studio, with the conditions set at 66 K and 1 bar (this environmental parameter was used for all subsequent calculations of hydrogen storage performance). The Dreiding force field was selected. The van der Waals interactions between molecules were described using the Lennard-Jones (L-J) 12-6 potential energy model, while hydrogen bonding interactions were characterized by the L-J 12-10 potential energy model. To accurately estimate electrostatic interactions, the Ewald summation method was employed as an effective means. During the simulation, non-covalent interactions between molecules and charge interactions between atoms were represented by the L-J 12-6 potential energy function and charge interactions, respectively. The calculation of potential energy parameters followed the Lorentz-Berthelot mixing rules.

In this study, Zr ions in the frameworks of UIO-66, UIO-67, UIO-68, MOF-801, MOF-802, and MOF-841 were substituted with Fe, Co, Ni, Cu, and Zn ions. To ensure the uniformity and comparability between the framework structures and the substituted metal ions during model construction, all Zr ions in UIO-66, UIO-67, UIO-68, MOF-801, MOF-802, and MOF-841 were selected as the targets for substitution, thereby generating the desired target structural models. The Zr-based MOFs material models used in the calculations can be constructed using the Materials Studio software and optimized structurally through the Forcite module. After optimization, the models can be regarded as rigid frameworks. The Lennard-Jones (L-J) potential energy parameters of the framework atoms are selected from the L-J potential energy parameters of the respective atoms in the Dreiding force field. The potential energy parameters for Zr, Fe, Co, Ni, Cu, and Zn atoms are taken from the UFF force field.

## 3. Results and discussion

The calculation results of physical properties of UIO-66, UIO-67, UIO-68, MOF-801, MOF-802, and MOF-841 are shown in Table S1 and Figure 2. As a series of materials, from UIO-66 to UIO-68, with the increase in the length of linkers, the unit volume, specific surface area, and porosity all exhibit a monotonically increasing trend, while the density decreases monotonically[17-19]. This is similar to the changes observed in MOF-801, MOF-802, and MOF-841. As can be seen from the figure, with the increase in the number of benzene rings in UiO-66, UiO-67, and UiO-68, the increase in porosity subsequently leads to a decrease in their bulk modulus, shear modulus, and Young's modulus, further proving the reliability of the computational method used in this paper[20, 21]. The calculations indicate that UiO-66 and MOF-801 have the highest bulk moduli, reaching 18.37 GPa and 20.49 GPa respectively, while UiO-67 and UiO-68 have the lowest, with only 6.8 GPa and 4.32 GPa. UiO-66 has the highest shear modulus at 62.09 GPa, followed by MOF-801 at 28.03 GPa, and again, UiO-67 and UiO-68 have the lowest values, at merely 1.49 GPa and 1.36 GPa. MOF-801 has the highest Young's modulus at 60.003 GPa, followed by MOF-802 and UiO-66 at 50.491 GPa and 47.31 GPa respectively, while UiO-68 has the lowest at only 8.36 GPa. The structure with the highest Poisson's ratio is MOF-841, and the structures with the lowest are MOF-801 and UiO-66. Meanwhile, it was found that there is a relationship between the mechanical properties of the materials and their specific surface areas. As the specific surface area increases, the mechanical properties decrease, yet a larger specific surface area is associated with a higher Young's modulus. From the perspective of molecular structure, the introduction of benzene rings alters the crystal packing mode of the materials. Each benzene ring has a relatively large planar structure, and when the number of benzene rings increases, the steric hindrance between molecules increases. This makes it difficult for molecules to form a closely packed and ordered structure during the packing process, resulting in the formation of more voids within the material and an increase in porosity. In terms of intermolecular forces, the originally potentially strong intermolecular interactions (such as hydrogen bonds and van der Waals forces) are disrupted by the increase in benzene rings[22, 23]. For example, the large $\pi$-bonds of benzene rings may affect the electron cloud distribution of surrounding molecules, weakening the intermolecular hydrogen bonding. When subjected to force, due to the weakened intermolecular forces and the presence of numerous voids, relative sliding and separation between molecules occur more easily. The material is unable to effectively resist volume compression (bulk modulus), shear deformation (shear modulus), and tensile-compressive deformation (Young's modulus) caused by external forces, leading to a decline in these mechanical property indicators.

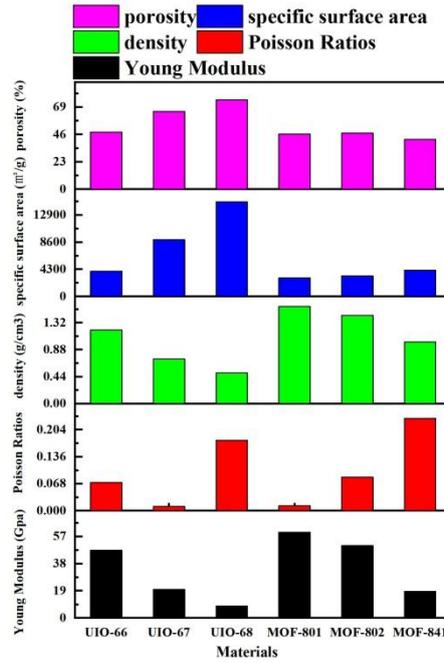

Figure 2 The calculation results of physical properties of UIO-66, UIO-67, UIO-68, MOF-801, MOF-802, and MOF-841.

The mechanical properties of MOF materials, such as strength and flexibility, directly influence their practical application performance. Taking the gas adsorption and separation process as an example, if the mechanical properties of the materials are inadequate, problems like structural collapse or pore deformation may occur, thereby reducing the adsorption efficiency. In catalytic reactions, MOF materials with insufficient mechanical stability may deactivate due to changes in reaction conditions. Thus, it can be seen that regulating mechanical properties is the key to enhancing the functional application effectiveness of MOF materials. By designing specific ligands, such as rigid or flexible ligands, the flexibility of the MOF framework can be regulated. Meanwhile, the coordination mode between metal ions and ligands, such as monodentate, bidentate, or polydentate coordination, also has a direct impact on structural stability. However, currently, the influence of central ion substitution on the mechanical properties of MOF materials has received relatively little attention. The substitution of different metal ions exerts varying impacts on the mechanical properties (Young's modulus and Poisson's ratio) of various MOF materials. The reasons may lie in the following aspects: On one hand, there are differences in the inherent properties of different metal ions. Those with larger atomic radii may render the structure of MOF materials looser. Different charge numbers result in different interaction forces between the metal ions and ligands, affecting the overall structural stability of the materials. On the other hand, the coordination mode (such as coordination number and coordination geometry) and coordination strength between metal ions and ligands vary. Complex coordination with high strength may make the material structure more compact, while the opposite may lead to a looser structure, thereby influencing the Young's modulus. Meanwhile, the coordination situation also affects the Poisson's ratio. In addition, different MOF materials inherently possess distinct structural characteristics, including topological structure, pore structure, and framework composition. These characteristics determine the extent and direction of the influence on the mechanical properties of the materials after metal ion substitution. For instance, for MOF materials with strong structural rigidity, metal ion substitution may have a relatively minor impact on their Young's modulus, while for those with good structural flexibility, it may significantly alter both their Young's modulus and Poisson's ratio. Specifically, for UIO-66, after substitution with Ni, Cu, and Zn, the Young's modulus increases and the Poisson's ratio decreases; after substitution with Fe and Co, the Young's modulus remains basically unchanged and the Poisson's ratio increases slightly. In the case of UIO-67, following substitution with Fe, Co, Ni, Cu, and Zn, both the Young's modulus and Poisson's ratio exhibit an upward trend, with Fe substitution resulting in the most significant increase. For UIO-68, Fe substitution leads to an increase in the Young's modulus. When substituted with Co, the Young's modulus remains largely stable. However, upon substitution with Ni, Cu, and Zn, it gradually declines, with Zn substitution causing the greatest decrease. In all these substitution scenarios, the Poisson's ratio increases, with Fe substitution yielding the smallest increment and Zn substitution bringing about the largest. Regarding MF-801, substitution with Fe and Cu causes the Young's modulus to rise. After substitution with Co and Ni, the Young's modulus stays essentially the same. But when substituted with Zn, it gradually drops. Across all these substitutions, the Poisson's ratio consistently increases, with Ni substitution leading to the smallest increase and Fe and Cu substitutions resulting in the largest. As for MOF-802, after substitution with Fe, Co, Ni, Cu, and Zn, the Young's modulus decreases in every instance, with Cu substitution causing the most substantial decrease. Meanwhile, the Poisson's ratio increases in all substitution cases, with MOF-802-Co and MOF-802-Ni having the largest Poisson's ratios. For MOF-841, substitution with Fe, Ni, and Zn leads to an increase in the Young's modulus. Conversely, substitution with Co and Cu causes it to decrease, with MOF-841-Zn having the largest value among them. Among the Poisson's ratios, those after substitution with Fe, Co, Ni, and Cu increase, whereas the one after Zn substitution decreases. MOF-841-Cu has the largest Poisson's ratio, and it can be inferred that the smallest should be MOF-841-Zn, not MOF-801-Zn.

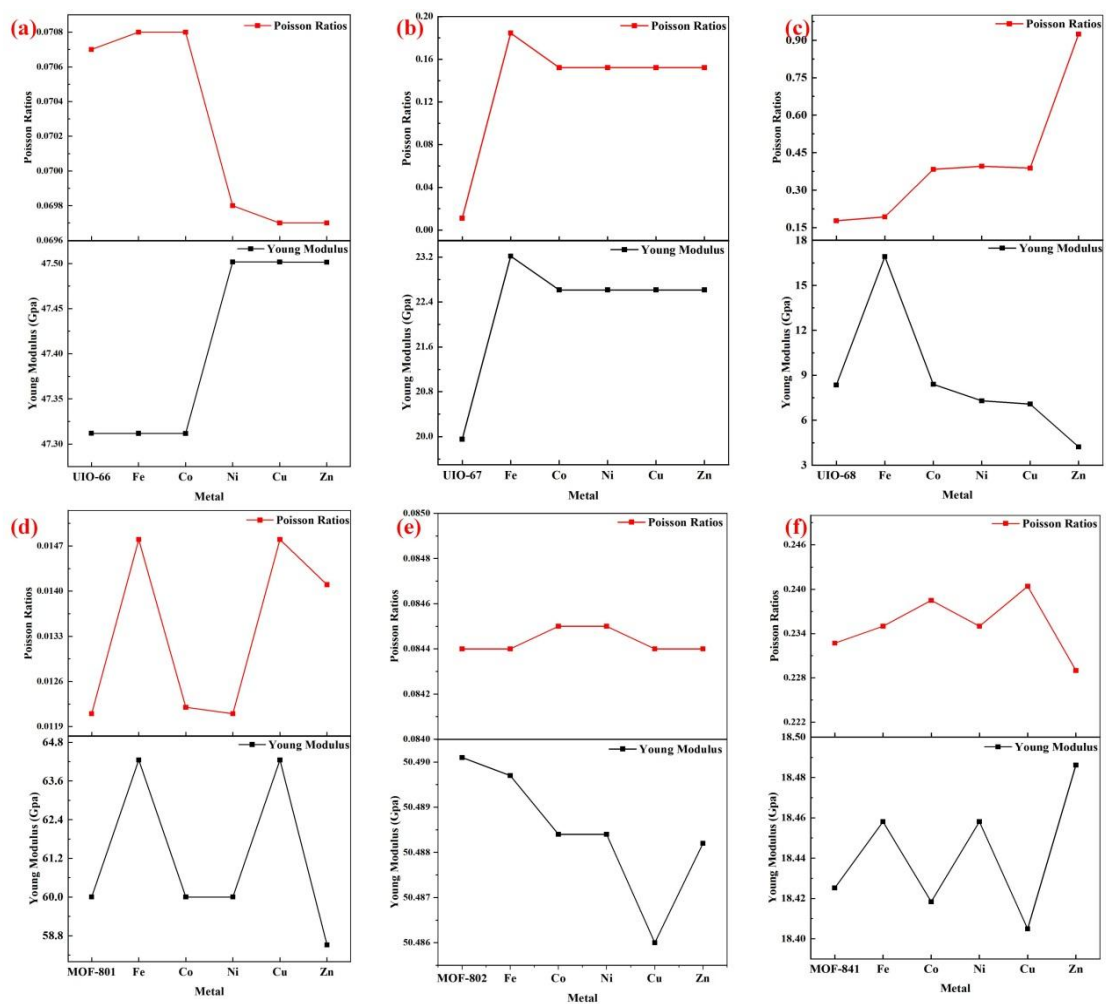

Figure 3 The substitution of different metal ions exerts distinct effects on the mechanical properties (Young's modulus and Poisson's ratio) of various MOF materials. (a) UIO-66, (b) UIO-67, (c) UIO-68, (d) MOF-801, (e) MOF-802, and (f) MOF-841.

Molecular simulation methods were employed to investigate the hydrogen storage performance of several Zr-based MOFs materials using the GCMC method. The adsorption capacity, heat of adsorption, and volumetric heat of adsorption of these Zr-based MOFs materials were compared under conditions of 66 K and 1 bar. Figure 4 illustrates the hydrogen adsorption capacity of Zr-based MOFs materials at 66 K. It can be observed that, in terms of the adsorption capacity of $H_2$ atoms per unit cell under conditions of 66 K and 1 bar, UiO-68 exhibits the highest adsorption capacity, followed by MOF-801, while MOF-841 has the lowest. The amount of adsorption is related to the unit cell size, porosity, and specific surface area of each crystal. Since the unit cell sizes, porosities, and specific surface areas of the aforementioned Zr-based MOFs materials vary, their unit cell adsorption capacities cannot be used as indicators of their hydrogen adsorption performance. As shown in the mass adsorption curves, UiO-68 has the highest volumetric adsorption capacity, reaching 6.53 STP cm$^3$/g, followed by UiO-67 with a mass adsorption capacity of 4.07 STP cm$^3$/g. MOF-802 has the lowest mass capacity, at only 0.188 STP cm$^3$/g. The hydrogen adsorption capacities of the other Zr-based MOFs materials are lower than those of UiO-67 and MOF-808, which is attributed to the high specific surface area and porosity of UiO-68. Conversely, MOF-802 has the smallest hydrogen storage capacity because it is constructed using H$_2$PZDC as a ligand, with the structure linked via pyrazole rings rather than benzene rings. Its low specific surface area and porosity limit its gas adsorption performance. In addition to the gas adsorption isotherms of materials under different pressures, the heat of adsorption ($Q$) is also an important parameter for evaluating the gas adsorption performance of materials. The magnitude of the heat of adsorption reflects the strength of the interaction between gas molecules and the framework structure. The $Q$ can be expressed by the following formula:

$$Q_{st} = \frac{\langle U \rangle \langle N \rangle - \langle UN \rangle}{\langle N^2 \rangle - \langle N \rangle \langle N \rangle} + RT \tag{1}$$

where, 〈 〉 represents the ensemble average, $\langle U \rangle$ represents the average energy of the structure, $\langle N \rangle$ represents the number of gas molecules, $R$ is the Boltzmann constant, and $T$ is the temperature of the system. As shown in the figure, the heat of adsorption curves of these different Zr-based MOFs materials at 66 K and 1 bar are compared. It can be seen that the differences in the heat of adsorption among the Zr-based MOFs materials with high adsorption heats are relatively small. The highest heat of adsorption is 0.209 Kcal/mol for MOF-801, with only a 1.95% and 12.3% difference from the second-highest

MOF-802 at 0.205 Kcal/mol and the third-highest UiO-66 at 0.186 Kcal/mol, respectively, and a 57% difference from the lowest heat of adsorption of UiO-68 at 0.133 Kcal/mol. A comparative study reveals a positive correlation between the hydrogen adsorption heat and Young's modulus of Zr-based MOFs materials. Specifically, the material with the highest adsorption heat also exhibits the highest Young's modulus, while the one with the lowest adsorption heat has the lowest Young's modulus. The ranking of adsorption heat values aligns consistently with that of Young's modulus values.

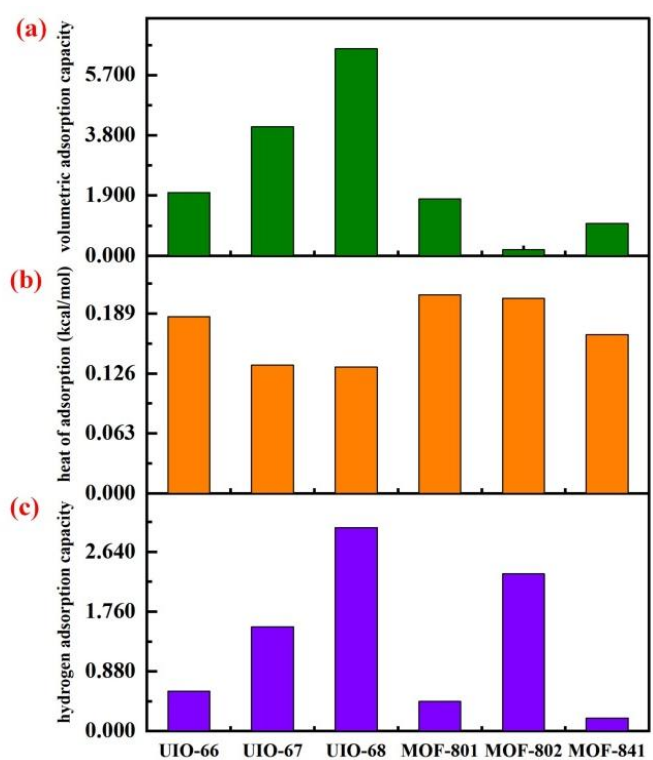

Figure 4 Hydrogen Storage Performance of Zr-Based MOFs Materials. (a) Volumetric Adsorption Capacity; (b) Adsorption Heat; (c) Hydrogen Adsorption Capacity.

On the basis of the existing structures of UiO-66, UiO-67, UiO-68, MOF-1, MOF-2, and MOF-41, a model construction approach was employed by substituting the Zr ions within their frameworks with Fe, Co, Ni, Cu, and Zn. This substitution of central metal ions was carried out to ensure consistency and comparability in the replacement of metal ions across the frameworks. To achieve this, all Zr ions in UiO-66, UiO-67, UiO-68, MOF-1, MOF-2, and MOF-41 were targeted for substitution, resulting in the desired structural models. The physical properties of the substituted structures are presented in Tables S2-S7. After the substitution of central metal ions, the free volume of the materials remained unchanged, and consequently, the porosity did not alter either. Compared to the original structures, the specific surface areas of the structures substituted with Fe, Co, Ni, Cu, and Zn all decreased. This reduction can be attributed to the differences in the relative atomic masses of these five metal elements. Specifically, the relative atomic masses of Fe, Co, Ni, Cu, and Zn are 55.85, 58.93, 58.69, 63.55, and 65.38, respectively, with Zr having the highest relative atomic mass among them and Fe the lowest. As a result, structures such as UIO-66-Fe, UIO-67-Fe, UIO-68-Fe, MOF-801-Fe, MOF-802-Fe, and MOF-841-Fe exhibited lower densities and larger specific surface areas. Conversely, structures like UIO-66-Zn, UIO-67-Zn, UIO-68-Zn, MOF-801-Zn, MOF-802-Zn, and MOF-841-Zn had higher densities and smaller specific surface areas. Among the materials substituted with metal ions, the sizes of the substituted metal ions followed the order Fe > Ni > Co > Cu > Zn. Therefore, substituting with lighter metal elements can increase the specific surface area of MOFs materials.

Figure 5 a presents the hydrogen storage performance of UIO-66 after substitution with Fe, Co, Ni, Cu, and Zn. As can be seen from the figure, the hydrogen adsorption capacity (Q) of UIO-66's unit cell is 0.59. The adsorption capacities of the materials substituted with Fe, Co, Ni, Cu, and Zn all decrease, being 0.549, 0.536, 0.541, 0.517, and 0.527, respectively. The order of adsorption capacities from highest to lowest is UIO-66-Fe > UIO-66-Ni > UIO-66-Co > UIO-66-Zn > UIO-66-Cu. The adsorption heat of UIO-66 is 0.186 kcal/mol, and the adsorption heats of the materials substituted with Fe, Co, Ni, Cu, and Zn all decrease, being 0.168 kcal/mol, 0.167 kcal/mol, 0.168 kcal/mol, 0.160 kcal/mol, and 0.170 kcal/mol, respectively. The order of adsorption heats from highest to lowest is UIO-66-Zn > UIO-66-Fe > UIO-66-Ni > UIO-66-Co > UIO-66-Cu. However, the order of volumetric adsorption capacities differs from the above. The volumetric adsorption capacity of UIO-66 is 1.99 STP cm³/g, and the volumetric adsorption capacities of the materials substituted with Fe, Co, Ni, Cu, and Zn all increase, being 2.19 STP cm³/g, 2.11 STP cm³/g, 2.13 STP cm³/g, 2.00 STP cm³/g, and 2.02 STP cm³/g, respectively. UIO-66-Fe has the highest volumetric adsorption capacity, which is 10.36% higher than that of UIO-66. The order of volumetric adsorption capacities from highest to lowest is UIO-66-Fe > UIO-66-Ni > UIO-66-Co > UIO-66-Zn > UIO-66-Cu. Obviously, metal ion substitution can increase both the elastic modulus and the volumetric hydrogen adsorption capacity of UIO-66. The hydrogen adsorption performances of samples after substitution with Fe, Co, Ni, Cu, and Zn are shown in Figure 5b-f. For UIO-67, after substituting the central metal ions with Fe, Co, Ni, Cu, and Zn, both the Young's modulus and volumetric adsorption capacity increase, with the samples substituted by Fe and Co showing the best performance. In contrast, for UIO-68, substitution with Fe, Co, Ni, Cu, and Zn does not significantly improve its mechanical properties and hydrogen storage performance. For MOF-801, substituting its central metal ions with Fe, Co, and Ni is feasible, with Fe being the optimal choice. For MOF-802, the substitution of its central metal ion with Fe represents the most favorable approach. And for MOF-841, its central metal ions can be replaced by Fe, Ni, and Zn, among which the substitution with Zn is the most superior option.

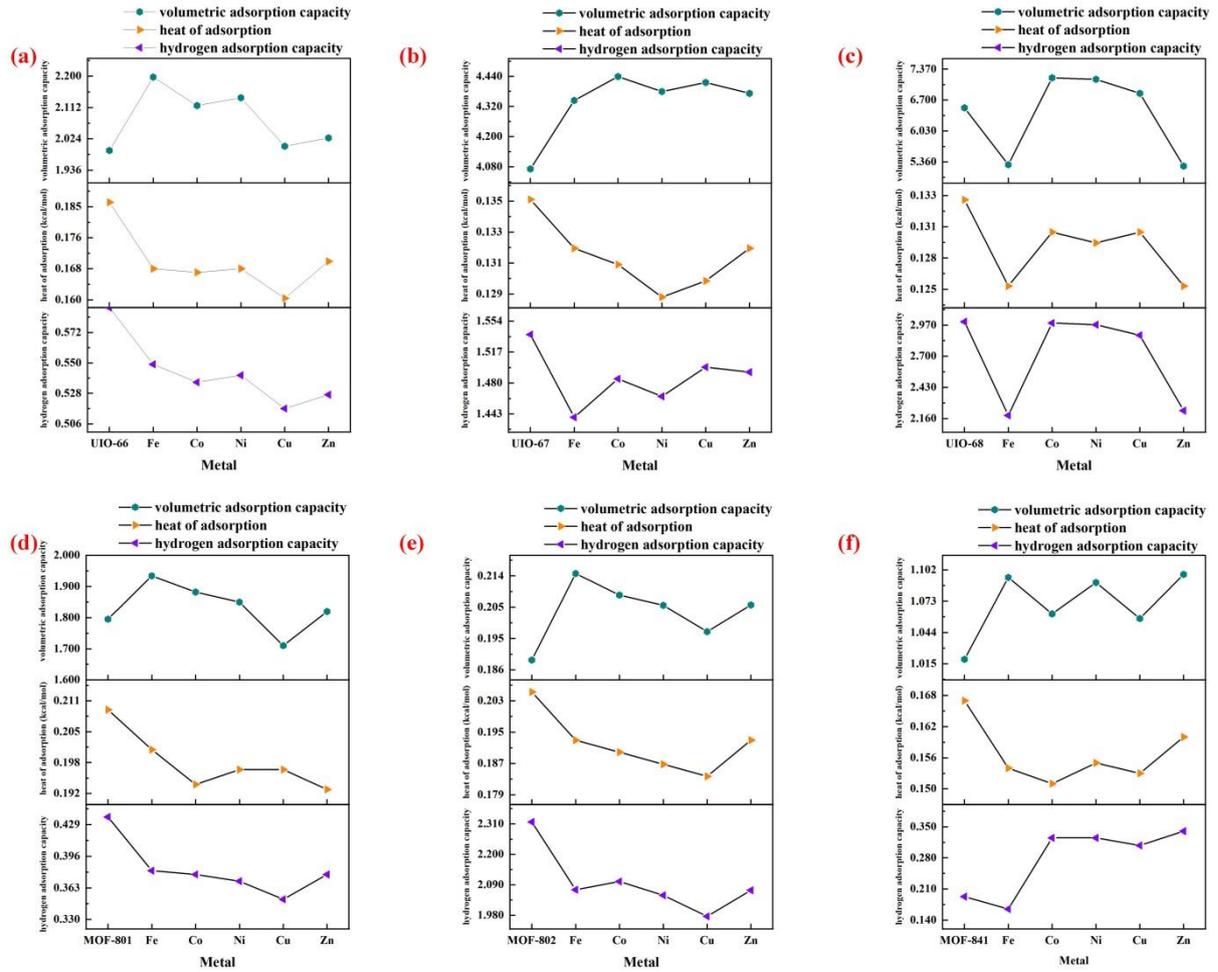

Figure 5 Hydrogen storage performance (volumetric adsorption capacity, adsorption heat, and hydrogen adsorption capacity) of various MOF materials substituted with Fe, Co, Ni, Cu, and Zn. (a) UIO-66; (b) UIO-67; (c) UIO-68; (d) MOF-801; (e) MOF-802; (f) MOF-841.

## 4. Conclusions

In this study, we conducted a comprehensive investigation into the mechanical and hydrogen storage properties of Zr-based MOFs under doping regulation. Through systematic computational analyses, we observed that substituting Zr ions with Fe, Co, Ni, Cu, and Zn ions in the frameworks of UIO-66, UIO-67, UIO-68, MOF-801, MOF-802, and MOF-841 leads to notable variations in their physical properties. Specifically, metal ion substitution affects the bulk modulus, shear modulus, Young's modulus, and Poisson's ratio of these MOFs, with different metal ions exerting distinct impacts. For instance, substituting Zr with Fe in UIO-66 enhances its Young's modulus and volumetric hydrogen adsorption capacity, while other substitutions may either increase or decrease these properties depending on the specific MOF material. Furthermore, our study reveals a positive correlation between the hydrogen adsorption heat and Young's modulus of Zr-based MOFs, indicating that materials with higher mechanical stability tend to exhibit better hydrogen storage performance. These findings underscore the importance of metal ion selection in the design of MOF materials for specific applications, such as hydrogen storage and catalysis.

**Acknowledgements**

This work was supported by Natural Science Foundation of Hunan Province (2025JJ70094).

**Competing interests**

The authors declare that they have no known competing financial interests or personal relationships that could have appeared to influence the work reported in this paper.

**Data Availability**

The datasets used and/or analyzed during the current study are available from the corresponding authors on reasonable request.

**Authors Contributions**

Dan Qian, and Zhipeng Liu: Investigation, Writing-original draft; Yanhuai Ding: Writing-review & editing, Project administration.

Transformation Journey of Zr-based MOFs: Study on Mechanics and Hydrogen Storage under Doping Regulation

Yanhuai Ding[*], Dan Qian, Zhipeng Liu

School of Mechanical Engineering and Mechanics, Xiangtan University, Xiangtan 411105, China

[*]Corresponding author. E-mail: yhding@xtu.edu.cn

Table S1 Calculation results of physical properties of raw Zr-based MOFs

|  | Density (g/cm$^3$) | Volume (Å$^3$) | Specific surface area (m$^2$/g) | Porosity (%) |
|---|---|---|---|---|
| UIO-66 | 1.198 | 9201.5 | 3991.6 | 47.831 |
| UIO-67 | 0.725 | 19370.6 | 8998.3 | 65.307 |
| UIO-68 | 0.499 | 34246.1 | 15016.2 | 74.942 |
| MOF-801 | 1.582 | 5724.8 | 2929.7 | 46.353 |
| MOF-802 | 1.439 | 28044.9 | 3268.2 | 47.055 |
| MOF-841 | 1.002 | 6031.9 | 4165.1 | 41.755 |

Table S2 Physical properties of UIO-66 after doping

|  | Density (g/cm$^3$) | Volume (Å$^3$) | Specific surface area (m$^2$/g) |
|---|---|---|---|
| UIO-66 | 1.198 | 9201.5 | 3991.6 |
| UIO-66-Fe | 1.045 | 9201.5 | 4576.8 |
| UIO-66-Co | 1.058 | 9201.5 | 4519.1 |
| UIO-66-Ni | 1.078 | 9201.5 | 4435.3 |
| UIO-66-Cu | 1.057 | 9201.5 | 4523.1 |
| UIO-66-Zn | 1.086 | 9201.5 | 4402.9 |

Table S3 Physical properties of UIO-67 after doping

|  | Density (g/cm$^3$) | Volume (Å$^3$) | Specific surface area (m$^2$/g) |
|---|---|---|---|
| UIO-67 | 0.725 | 19370.6 | 8998.3 |
| UIO-67-Fe | 0.653 | 19370.6 | 10001.1 |
| UIO-67-Co | 0.659 | 19370.6 | 9904.8 |
| UIO-67-Ni | 0.658 | 19370.6 | 9911.7 |
| UIO-67-Cu | 0.668 | 19370.6 | 9764.3 |
| UIO-67-Zn | 0.672 | 19370.6 | 9709.5 |

Table S4 Physical properties of UIO-68 after doping

|  | Density (g/cm$^3$) | Volume (Å$^3$) | Specific surface area (m$^2$/g) |
|---|---|---|---|
| **UIO-68** | 0.499 | 34246.1 | 15016.2 |
| **UIO-68-Fe** | 0.457 | 34246.1 | 16366.1 |
| **UIO-68-Co** | 0.461 | 34246.1 | 16238.7 |
| **UIO-68-Ni** | 0.461 | 34246.1 | 16247.9 |
| **UIO-68-Cu** | 0.466 | 34246.1 | 16052.1 |
| **UIO-68-Zn** | 0.469 | 34246.1 | 15979.1 |

Table S5 Physical properties of MOF-801 after doping

|  | Density (g/cm$^3$) | Volume (Å$^3$) | Specific surface area (m$^2$/g) |
|---|---|---|---|
| **MOF-801** | 1.582 | 5724.8 | 2929.7 |
| **MOF-801-Fe** | 1.335 | 5724.8 | 3469.7 |
| **MOF-801-Co** | 1.357 | 5724.8 | 3414.8 |
| **MOF-801-Ni** | 1.355 | 5724.8 | 3418.7 |
| **MOF-801-Cu** | 1.389 | 5724.8 | 3335.9 |
| **MOF-801-Zn** | 1.402 | 5724.8 | 3305.5 |

Table S6 Physical properties of MOF-802 after doping

|  | Density (g/cm$^3$) | Volume (Å$^3$) | Specific surface area (m$^2$/g) |
|---|---|---|---|
| **MOF-802** | 1.439 | 28044.9 | 3268.2 |
| **MOF-802-Fe** | 1.238 | 28044.9 | 3798.6 |
| **MOF-802-Co** | 1.256 | 28044.9 | 3745.6 |
| **MOF-802-Ni** | 1.255 | 28044.9 | 3749.4 |
| **MOF-802-Cu** | 1.282 | 28044.9 | 3669.0 |
| **MOF-802-Zn** | 1.292 | 28044.9 | 3639.5 |

Table S7 Physical properties of MOF-841 after doping

|            | Density (g/cm³) | Volume (Å³) | Specific surface area (m²/g) |
|---|---|---|---|
| **MOF-841**    | 1.002 | 6031.9 | 4165.1 |
| **MOF-841-Fe** | 0.885 | 6031.9 | 4714.5 |
| **MOF-841-Co** | 0.895 | 6031.9 | 4660.8 |
| **MOF-841-Ni** | 0.895 | 6031.9 | 4664.7 |
| **MOF-841-Cu** | 0.911 | 6031.9 | 4582.9 |
| **MOF-841-Zn** | 0.917 | 6031.9 | 4552.6 |